\documentclass[twocolumn,showpacs]{revtex4}
\usepackage{graphicx}% Include figure files
\usepackage{dcolumn}% Align table columns on decimal point
\input epsf
\usepackage{graphicx}
\def\be{\begin{equation}}
\def\ee{\end{equation}}
\def\bea{\begin{eqnarray}}
\def\eea{\end{eqnarray}}

\begin{document}
 \tighten
%\input epsf
%%\draft
\renewcommand{\topfraction}{0.8}
\vskip 3cm

\

\preprint{SU-ITP-04/22}
\title {{\Large\bf Deformation, non-commutativity and \\ \hskip 1cm \vskip -7pt
the cosmological constant
problem\footnote{A talk at the  conference in honor of Albert
Schwarz. The PowerPoint version of this talk can be
found, using Microsoft Explorer, at http://www.stanford.edu/people/kallosh/Talks/Albert.pps
}}}

 \author{\bf   Renata Kallosh }
\address{ {Department
  of Physics, Stanford University, Stanford, CA 94305-4060,
USA}    }

{\begin{abstract} In this talk we provide arguments  on
possible relation between the cosmological constant in our space
and the non-commutativity parameter  of the internal space of
compactified string theory. The arguments are valid in the context
of  D3/D7 brane cosmological model of inflation/acceleration.
\end{abstract}}
\maketitle
%\tableofcontents{}

The current set of cosmological data strongly suggests that the
universe experienced at least two separate stages of inflation,
which is an accelerated expansion of the universe (${\ddot a}>0$)
in a de Sitter-like state. The first inflationary stage began 13.7
billion years ago, immediately after the Big Bang. It could be
very short, lasting only approximately $10^{-35}$ seconds, but it
was long enough to make our universe extremely large and to
produce density perturbations responsible for the formation of
galaxies. The initial value of the Hubble constant $H$ in the
beginning of inflation could be arbitrarily large, but its value
at the last stages of inflation should be several orders of
magnitude smaller than the Planck mass, $H_{\rm infl}\leq
10^{-5}M_{P}$, since otherwise we would see gravitational waves
produced during inflation. At this stage the vacuum energy was
given by $V\sim H^2 M_{P}^2 < 10^{-10} M_P^4$.

After inflation there was a long stage when  the universe kept
expanding, ${\dot a}>0$, but it was decelerating, ${\ddot a}<0$.
But later, approximately 5 billion years ago, the second stage of
accelerated expansion began, with the  Hubble parameter
$H_{accel}\sim 10^{-60}M_{P}$. At this stage the vacuum energy is
given by $V\sim H^2 M_{P}^2\sim 10^{-120} M_{P}^4$; it constitutes
about 70\% of the total energy density of the universe at present.
Thus, string theory faces a challenge to explain  these two
stages of accelerated expansion of the universe.

String theory and cosmology is an emergent topic. All observations
so far fit 4-dimensional Einstein general relativity. The major
problem here is how to get this picture and the explanation of the
data from the compactified fundamental 10-dimensional string
theory or 11-dimensional M-theory and supergravity. First of all,
we should learn how to get any 4-dimensional de Sitter space from
some string theory construction. Secondly, we would like to be
more ambitious and explain the numbers like $H_{\rm infl}\leq
10^{-5}M_{P}$ and $H_{\rm accel}\sim 10^{-60}M_{P}$.

Over the last few years there were several developments in string
theory  which attempted to address the challenge. In type-IIB
string theory Giddings, Kachru and Polchinski \cite{GKP} have
found a way to stabilize the dilaton-axion moduli using the 3-form
fluxes. Also the proposal was made about the possibility
of volume stabilization mechanism by Kachru, R. K.,  Linde and
Trivedi, the KKLT model of de Sitter space \cite{KKLT}. New
interesting developments  in this direction with the name of
``Landscape of string theory'' were proposed by Susskind
\cite{Susskind:2003kw} and ``Statistics of flux vacua'' by Douglas
\cite{Douglas:2004kp} in the spirit of  Bousso-Polchinski \cite{Bousso:2000xa} proposal to tune the cosmological constant using the large number of flux vacua.
Explicit string models with stabilized
moduli have been constructed 
\cite{Denef:2004dm}.  KKLT model of de Sitter space was developed into a model of
inflation by  Kachru, R. K., Maldacena, McAllister, Linde and
Trivedi, the \mbox{K\hspace{-5.pt}KLM\hspace{-7.pt}MT} model \cite{KKLMMT}. The picture describing this model (as
well as the \mbox{K\hspace{-5.pt}KLM\hspace{-7.pt}MT} symbol) is
taken from 
\cite{Copeland:2003bj}:
\begin{figure}[h!]
\centering\leavevmode\epsfysize=4cm \epsfbox{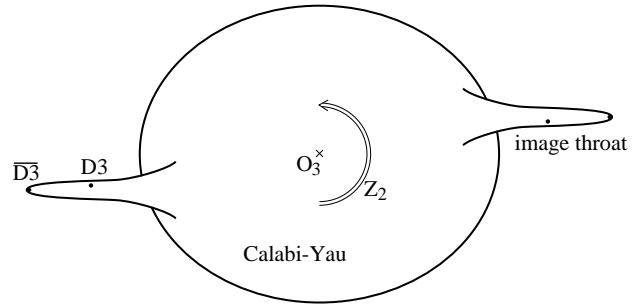}
\caption[fig1] {Schematic picture of the
\mbox{K\hspace{-5.pt}KLM\hspace{-7.pt}MT}  geometry: a warped
Calabi-Yau manifold with throats, identified under a ${\bf Z}_2$
orientifold.} \label{kklt}
\end{figure}
The throat geometry has a highly warped region related to the tip
of the resolved conifold:
$$w_1^2+w_2^2+w^2_3+w_4^2=z\ ,$$ where the warping factor  $e^{2A}$ in
$ds^2= e^{2A(y)} ds^2_4 + ds_y^2$
reaches its minimal value:
$$e^{A_{min}}\sim z^{1/3} \sim e^{-{2\pi K\over 3Mg_s}}\qquad e^{2A_{min}}\ll 1 \ .$$
Warped geometry of the compactified space and nonperturbative
effects  allow to obtain AdS space with unbroken SUSY and
stabilized volume. One can uplift AdS space to a metastable dS
space by adding anti-D3 brane at the tip of the conifold.
\begin{figure}[h!]
\centering\leavevmode\epsfysize=4cm \epsfbox{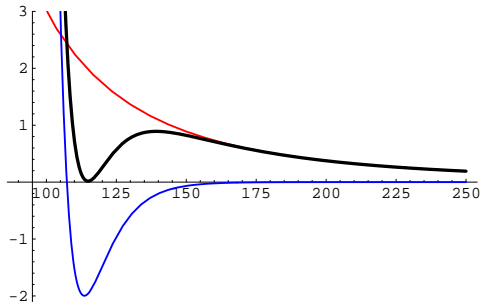}
\caption[fig1] {Uplifting of the AdS (blue) minimum at negative
energy to dS (black)  minimum at positive energy due to the term
in the potential (red curve) which has a runaway behavior towards
large volumes.} \label{KKLTpotential}
\end{figure}
The small warping factor plays a significant role in the uplifting
AdS  vacuum to dS. The smallness of $ z $ is due to the resolution
of the conifold singularity. The KKLT potential  consists of two
parts. The $V_F$ part comes from the non-perturbative
superpotential and permits an AdS vacuum (the blue plot with the
minimum at negative energy). The second contribution proportional
to $\sigma^{-3}$ (the red runaway curve) comes from anti-D3 brane,
placed at the tip of the throat
$$V= V_F + V({\bar D3}) \qquad V({\bar D3})=
 {C\over \sigma^3} \ . $$
Here $ C\sim z^{2/3}\sim e^{4A_{min}}\ll 1.$ In our example $C$
was $10^{-9}$. Small  $C$ is necessary for dialing  the anti-D3
energy to AdS scale to preserve and uplift the minimum to the
position with a small positive value of $V$, as shown in Fig.
\ref{KKLTpotential} (in the black curve). The redshift in the
throat plays the key role in this construction.

Thus, the advantage of using highly warped geometry is that we
have a source of small parameters in string theory, where
otherwise we have only the  string scale and discrete fluxes and
{\it a priory} no small parameters.

The disadvantage of relying totally on high warping shows up in
the  basic version of the inflationary
\mbox{K\hspace{-5.pt}KLM\hspace{-7.pt}MT} model: highly warped
region of Klebanov-Strassler  geometry \cite{Klebanov:2000hb}
corresponds to conformal coupling of the inflaton field (position
of D3-brane in the throat region) which leads to
$$M^2_{{}_{\rm infl}}\sim H^2 \ .$$
However, the observed flatness of the spectrum of inflationary perturbations requires that
$$M^2_{{}_{\rm infl}} \sim 10^{-2} H^2 \ .$$
Few possibilities to improve the basic
\mbox{K\hspace{-5.pt}KLM\hspace{-7.pt}MT}  model are already
known. Here, however, we will proceed with an alternative
scenario, initiated by the recent work of Burgess, R. K. and
Quevedo \cite{Burgess:2003ic},  where the role of uplifting of AdS
to dS was given not to an anti-D3 brane, but to a D7 brane with
2-form fluxes on its world-volume, which also lead to the term
proportional to $\sigma^{-3}$. Here again, to get a dS
stabilization of the volume we had to place D7 in the highly
redshifted region of the moduli space.

{\it The main proposal of this talk is to switch to a new source
of  uplifting which is based on a non-commutative nature of the
space, orthogonal to D3 brane in the D3/D7 brane system}.

The D3/D7 cosmological model proposed by Dasgupta, Herdeiro,
Hirano and R. K.  a while ago \cite{DHHK}, before the issue of
volume stabilization was developed, has been recently
reconsidered. This model gives a stringy realization of Linde's
hybrid inflation model \cite{Hybrid},  in particular, of its
version related to the D-term inflation model of Binetruy-Dvali
and Halyo \cite{Binetruy:1996xj}. This follows the general setting
of brane inflation concept by Dvali and Tye \cite{Dvali:1998pa}.

\begin{figure}[h!]
\centering\leavevmode\epsfysize=4cm \epsfbox{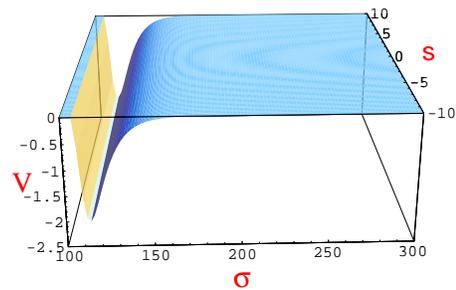}
\caption[fig1] {AdS trench, the exact supersymmetric state
corresponds to the minimum of the potential in the
$\sigma$-volume-direction and shows the neutral equilibrium in the
$s$-inflaton-direction.} \label{AdS}
\end{figure}

\begin{figure}[h!]
\centering\leavevmode\epsfysize=4cm \epsfbox{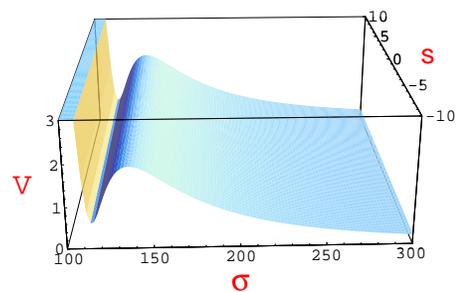}
\caption[fig1] {dS trench at the minimum of  the volume $\sigma$
is stabilized close to  the critical value of the AdS stage. The
inflaton dependence is still nearly flat.} \label{dS}
\end{figure}

In the subsequent work with Prokushkin and Hsu
\cite{KalloshHsuProk}, we have combined the original D3/D7 model
with KKLT construction providing the volume stabilization. The
corresponding mechanism, which we called ``inflaton trench,''
shows that one can get an extra dimension to both models. Both
pictures display the concept of the inflaton shift symmetry: in
certain situations the motion of branes does not destabilize the
volume, which was also discussed by Firouzjahi and Tye
\cite{Firouzjahi:2003zy}. The proposed inflaton shift symmetry of
this model was confirmed more recently by Hsu and R. K.
\cite{Hsu:2004hi} using an important input from the work of
Angelantonj, D'Auria, Ferrara and Trigiante
\cite{Angelantonj:2003up} as well as Koyama, Tachikawa and Watari
\cite{KTW} describing this model in the context of special
geometry. In familiar case of near extremal black holes duality
symmetry protects exact entropy formula from large quantum
corrections. In the cosmological context duality (inflaton shift
symmetry) protects  the flatness of the potential in D3/D7
inflation model from large quantum corrections.
\begin{figure}[h!]
\centering\leavevmode\epsfysize=6 cm \epsfbox{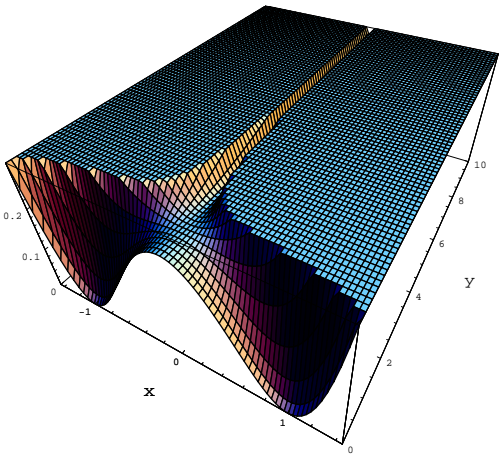}
\caption[fig1] {The hybrid inflation potential of the D3/D7 model.
} \label{1c}
\end{figure}
\begin{equation}V= S^2 \Phi^\dagger \Phi + {g^2\over 2} D^2\, , \qquad \vec D= \Phi^\dagger \vec \sigma \Phi - \vec \xi \ .\end{equation} 
Here $\Phi$ is  the charged hypermultiplet and $\vec \xi$ is an FI
triplet, providing the resolution of the small instanton
singularity, $S$ is the distance between branes.

\begin{figure}[h!]
\begin{picture}(75,0)(0,0)
 \put(-35,-100){$D7$}
\put(-10,-145){$x^{4,5}$} \put(80,-100){$D3$}
 \put(-90,-120){$x^{0,1,2,3}$} \put(-20,-120){$x^{6,7,8,9}$}
 %\put(100,90){$_{\sigma=\pi}$} \put(245,112){$_{\sigma=0}$}
 \put(-35,-32){${\mathcal F}$}
\end{picture}
\centering \epsfysize=13cm
\includegraphics[scale=0.4]{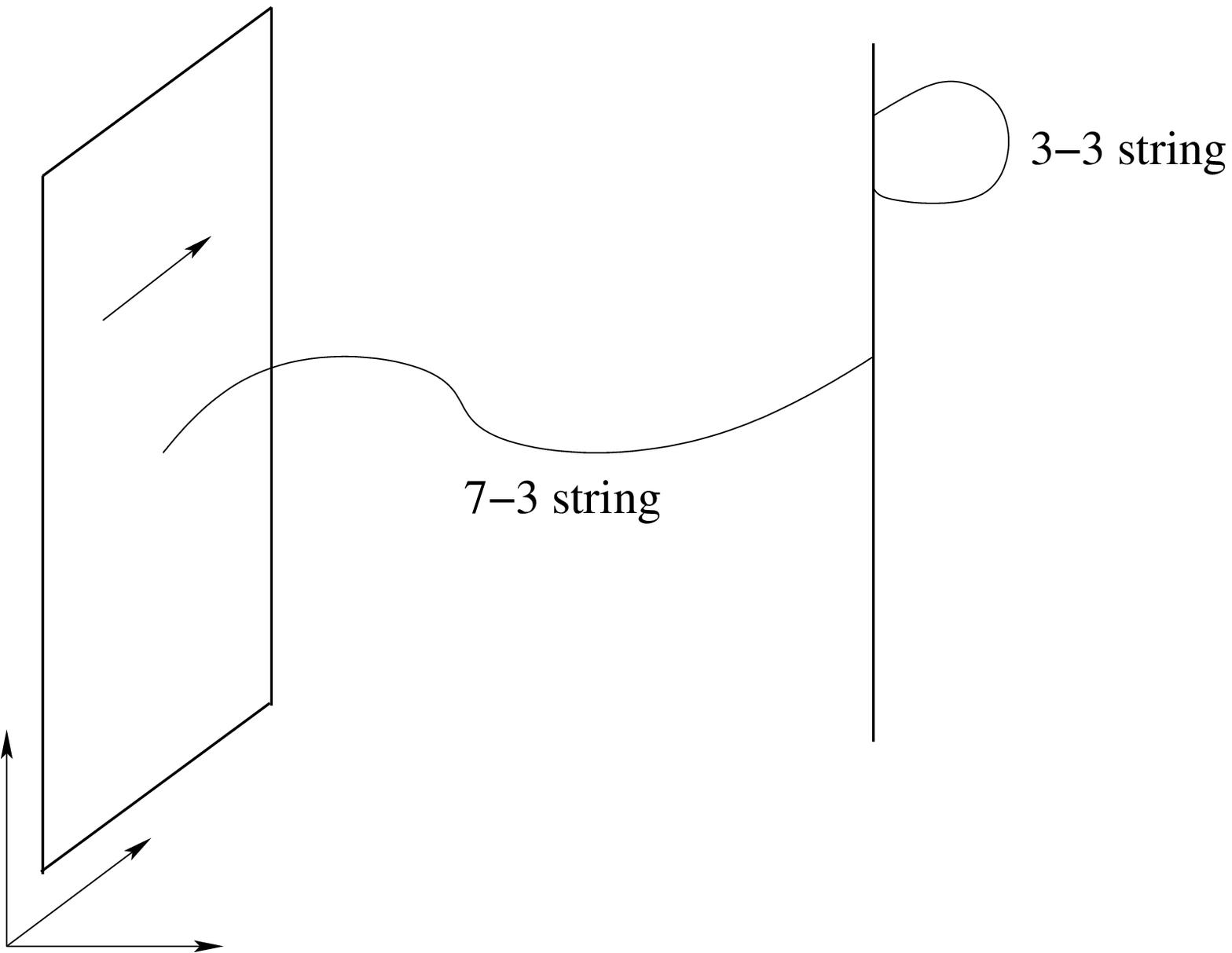}
%{branehybr1.eps}
\caption{The D3/D7 ``cosmological" system. The 3-3 strings give
rise to the ${\mathcal N}=2$ vector multiplet, the 7-3 strings to
the hypermultiplet and the worldvolume 2-form field ${\mathcal F}= dA-B$
to the FI terms of the $D=4$ gauge theory.}
\end{figure}

The mass of D3-D7 strings (hypers) is split due to the presence of
the deformed flux on D7. This leads to the one-loop effect
correction to the potential:
\begin{equation}V_{1-{\rm loop}}= {g^2\xi^2\over 2}[1+ {g^2\over 8\pi^2} \ln
{S^2\over S^2_{cr}}]\  .\end{equation} Here $S_{cr}$ is the critical value of the
inflaton field where dS minimum  turns into dS maximum.
De Sitter stage of the model (the long uplifted valley in Fig. \ref{1c}):
$$\Phi=0\ , \qquad S\gg S_{cr}\ , \qquad  \vec D= \vec \xi \ ,
\qquad V\approx {g^2\xi^2\over 2}$$
can explain inflation or current acceleration, depending on the parameters of the model.
The D3 brane is attracted to the D7 brane due to the presence of
the  anti-self-dual deformed flux on D7 in accordance with spontaneously
broken supersymmetry in de Sitter valley. When the critical
distance between branes has been reached, the waterfall stage
brings the system into the supersymmetric ground state:
$$\Phi^\dagger  \vec \sigma \Phi=\vec \xi\ , \qquad S=0 \ ,
\qquad
 \vec D= 0\ ,
\qquad V= 0 \ .$$ D3/D7 bound state corresponds to the Higgs
branch and non-commutative instantons.  It is described by
Nekrasov-Schwarz non-commutative instantons
\cite{Nekrasov:1998ss}.
\begin{figure}[h!]
\centering\leavevmode\epsfysize=6 cm \epsfbox{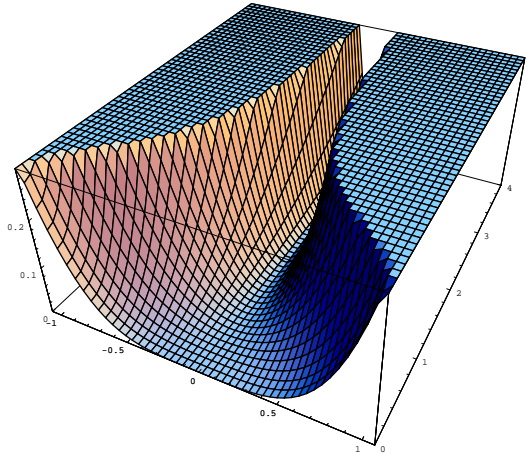}
\caption[fig1] {The potential of the D3/D7 without
Fayet-Iliopoulos term, $\vec \xi=0$.} \label{nofi}
\end{figure}
D3 can move away from D7 when the deformation parameter $\vec \xi$
vanishes,  the moduli space is singular: there is no de Sitter
space as shown in Fig. \ref{nofi}.

Resolution of singularity of the moduli space of instantons in
D3/D7 Higgs branch requires $\vec \xi\neq 0$ so that the Coulomb
branch $\vec D=\Phi^\dagger \vec \sigma \Phi -\vec \xi$  with
$\Phi =0$ has a non-vanishing D-term potential $V_D={g^2
\xi^2\over 2}$. Thus we see the relation between the deformation,
non-commutativity and resolution of singularity which leads to de
Sitter space in effective 4-dimensional space. The relation
between Nekrasov-Schwarz non-commutative instantons and non-linear
deformed instantons in Dirac-Born-Infeld theory was explained by
Seiberg and Witten \cite{seibergwittenncg}.

To be more specific, in the brane construction we consider
Dirac-Born-Infeld $\kappa$-symmetric action and non-linear
deformed instantons of Marino, Minassian, Moore and Strominger
\cite{Marino:1999af} which were originally constructed for  D0/D4
system. The construction of a D3/D7 bound state with unbroken
supersymmetry is based on definition of unbroken supersymmetry of
a set of branes via the equation
$$(1-\Gamma)\epsilon=0 \ ,$$
where $\Gamma$ is the $\kappa$-symmetry operator on D7 brane in
the background of the D3 brane with some constant two-form $B$.
According to Bergshoeff, R. K., Ort\'{\i}n and  Papadopoulos
\cite{Bergshoeff:1997kr}, the complete dependence of $\Gamma$ of
the deformed 2-form flux on the world-volume, ${\cal F}= dA- B$,
is given by a ``rotating factor'' $a$
$$\Gamma= e^{-a/2} \Gamma_0 e^{a/2}\ ,$$
where $a$ depends on a 2-form $Y$
$$a={1\over 2} Y_{ik}\Gamma^{ik}\sigma_3\ ,$$
which in turn is a complicated (but known) non-linear function of ${\cal F}$
$${\cal F}= {\rm ``\tan"}\, Y \ .$$
This $\kappa$-symmetric construction is a generalization and
formalization of the work of Berkooz, Douglas and Leigh on
unbroken supersymmetry of a combination of branes intersecting at
angles \cite{Berkooz:1996km}. Non-linear deformed Abelian
instanton equation
$${{\cal F}^-\over 1+{\rm Pf}\, {\cal F}}=
- {B^-\over 1+{\rm Pf}\, B}$$ follows from $(1-\Gamma)\epsilon=0$,
it has finite energy due to non-vanishing deformation
(non-commutativity) parameter $B^-$. Here ${\cal F}^-$ is an
anti-self-dual part of the two-form  ${\cal F}$ and $B^-$ is an
anti-self-dual part of the two-form $B$.

Now we can to  go back to the D-term volume stabilization
issue:  instead of anti-D3 brane we add D7 brane.  The D-term
potential depends on the anti-self-dual deformed flux ${\cal F}^-$
and volume modulus
$$V_D \sim {(\theta_1-\theta_2)^2\over \sigma^3}\ ,$$
where $ \theta_1 = \arctan {\cal F}_{67}$ and
  $\theta_2 = \arctan {\cal F}_{89} $.
When $\theta_1$ and $\theta_2$ are small we have
$$V_D \sim {(\theta_1-\theta_2)^2\over \sigma^3}
\approx {({\cal F}^-)^2\over \sigma^3}\ .$$ There are 2 possibilities
to make this mechanism working for the  uplifting of AdS minimum
to dS minimum:
\\

A) Place D7 in highly warped region of space as proposed in
Burgess, R. K., Quevedo \cite{Burgess:2003ic}. The required small
fine-tuning parameter $C$ is due to the warping associated with
the deformation of the conifold $z$.

B)  Use non-commutativity of the internal space:  irrational $B$
cannot be gauged away into quantized $F=dA$. Quantization of the flux
$\int F=2\pi n$ may have prevented the D-term contribution from D7 of
the form ${C\over \sigma^3}$ to provide a small $C$ when D7 is not at
the tip of the conifold but somewhere in the part of Calabi-Yau space.
{\it Here the deformation parameter $B^-$ in ${\cal F}^-$ comes to rescue,
it is not quantized, it can be small! We may choose therefore the vanishing
quantized part of the 2-form so that ${\cal F}^-$ may be tuned via
deformation parameter to uplift any AdS minimum into a de Sitter
minimum since $C\sim (B^-)^2$. }

In the context of non-commutative  Nekrasov-Schwarz instantons
and of Seiberg-Witten  non-linear instantons  in DBI action, Fayet-Iliopoulos
terms  are necessary to make the Abelian  instantons non-singular.
It is tempting to speculate that in D3/D7 cosmological model with
volume  stabilization mechanism there is an explanation of the
non-vanishing effective cosmological constant: the
non-commutativity parameter (FI term in effective theory) is
needed to remove the instanton  moduli space singularity in the
description of the supersymmetric D3/D7 bound state when D3 is
dissolved into D7. The same cosmological model must have a
non-supersymmetric de Sitter stage when D3 is separated from D7.

Both methods A) and B) can be used for uplifting the AdS vacuum to dS vacuum, which is necessary for describing the present state of acceleration of the universe with $H_{accel}\sim 10^{-60}M_{P}$. However, the method B) based on the use of FI terms and non-commutativity seems to have an important advantage when we want to make the second uplifting of the vacuum energy in order to describe the stage of inflation: This method allows us to obtain in a natural way an inflationary flat direction with $m^2 \ll H^2$, as shown in Figs. 4 and 5.

It is quite interesting that the amplitude of inflationary  perturbations of metric and the resulting CMB anisotropy in the theory with the potential (1), (2)   coincides with the FI term $\xi$, up to a numerical coefficient $O(1)$ \cite{pterm}. This suggests an intriguing possibility that one can measure the non-commutativity parameters of the internal space
by looking at the sky.

\

It is a pleasure to thank T. Banks, A. Dabholkar, K. Dasgupta, M.
Douglas, B. Freivogel, S. Kachru,  A. Linde, J. Maldacena,  M. Rieffel, A.
Schwarz, J. Schwarz,  M. Sheikh-Jabbari and  L. Susskind,  for
most useful discussions.  This work was supported  by NSF grant
0244728.

\end{document}